\documentclass[10pt,conference]{IEEEtran}
\usepackage{graphicx} 
\usepackage{cite}

\title{OrQstrator: An AI-Powered Framework for Advanced Quantum Circuit Optimization}

\author{\IEEEauthorblockN{Laura Baird}
\IEEEauthorblockA{Department of Computer Science \\
University of Colorado Colorado Springs (UCCS), USA \\
lbaird@uccs.edu}
\and
\IEEEauthorblockN{Armin Moin}
\IEEEauthorblockA{Department of Computer Science \\
University of Colorado Colorado Springs (UCCS), USA \\
amoin@uccs.edu}
}

\begin{document}

\maketitle
\begin{abstract}
We propose a novel approach, OrQstrator, which is a modular framework for conducting quantum circuit optimization in the Noisy Intermediate-Scale Quantum (NISQ) era. Our framework is powered by Deep Reinforcement Learning (DRL). Our orchestration engine intelligently selects among three complementary circuit optimizers: A DRL-based circuit rewriter trained to reduce depth and gate count via learned rewrite sequences; a domain-specific optimizer that performs efficient local gate resynthesis and numeric optimization; a parameterized circuit instantiator that improves compilation by optimizing template circuits during gate set translation. These modules are coordinated by a central orchestration engine that learns coordination policies based on circuit structure, hardware constraints, and backend-aware performance features such as gate count, depth, and expected fidelity. The system outputs an optimized circuit for hardware-aware transpilation and execution, leveraging techniques from an existing state-of-the-art approach, called the NISQ Analyzer, to adapt to backend constraints. 
\end{abstract}


\section{Introduction}
As quantum computing advances at a remarkable pace, it is anticipated that fault-tolerant architectures could be realized sooner than previously expected \cite{bolgarMicrosoftsMajorana1}. Despite these breakthroughs, practical Fault-Tolerant Quantum Computing (FTQC) remains currently out of reach. In the meantime, we remain firmly in the Noisy Intermediate-Scale Quantum (NISQ) era \cite{preskillQuantumComputingNISQ2018}, an era defined by hardware limitations: limited qubit counts, short circuit coherence times, and high gate error rates.

To reliably run algorithms under these constraints, quantum circuits must be significantly optimized. This means reducing circuit depth, minimizing two-qubit gate usage, and adapting to the hardware-specific limitations of each of the heterogeneous quantum computing platforms. While optimization strategies, such as rule-based rewriting \cite{amyMeetintheMiddleAlgorithmFast2013, heyfronEfficientQuantumCompiler2018a}, domain-specific numeric methods \cite{kuklianskyQFactorDomainSpecificOptimizer2023, xuOptimizingQuantumCircuits2025}, and Deep Reinforcement Learning (DRL) \cite{foselQuantumCircuitOptimization2021, riuReinforcementLearningBased2025} have demonstrated their benefits, they were developed and evaluated in isolation, and only tackle one of the optimization problems at a time. Static compiler pipelines struggle to generalize across diverse circuit structures and backend architectures, limiting their effectiveness in real-world scenarios.

The contribution of this extended abstract accompanying our poster is to increase the effectiveness of the optimization approaches through an AI-powered orchestration framework that learns to coordinate a suite of complementary optimization techniques. By integrating learned DRL agents, parameterized instantiation strategies, and domain-specific numeric optimizers, the proposed framework, called OrQstrator, dynamically generates hardware-adapted, fidelity-aware quantum circuits suitable for near-term execution. In the following, we briefly review the literature. We then propose our novel approach, conclude, and suggest future work.

\section{Related Work}  
Fösel et al. \cite{foselQuantumCircuitOptimization2021} demonstrated that their DRL-based approach could learn powerful rewrite strategies, achieving up to 27\% circuit depth reduction. Moreover, Riu et al. \cite{riuReinforcementLearningBased2025} extended this by using ZX-calculus-based transformations combined with reinforcement learning to generalize to larger circuits. Further, Moro et al. \cite{moroQuantumCompilingDeep2021} applied DRL to quantum compiling, showing its utility for gate synthesis.

On the classical side, Kukliansky et al. \cite{kuklianskyQFactorDomainSpecificOptimizer2023} introduced QFactor, a domain-specific optimizer that used analytical gradients and tensor networks to perform logical gate resynthesis. Also, Younis and Iancu \cite{younisQuantumCircuitOptimization2022a} proposed a method for parameterized circuit instantiation that improved both optimization and gate set translation, yielding $\sim$13\% average gate count reduction. These methods, while powerful individually, were limited by their fixed pipelines and narrow scope.

Hardware-aware compilation has also received attention. Salm et al. \cite{dustdar_nisq_2020} developed the NISQ Analyzer, which would match quantum algorithms to feasible hardware targets by evaluating device-specific constraints. Other works (e.g., Murali et al. \cite{muraliNoiseAdaptiveCompilerMappings2019a}) have explored routing and mapping using noise-aware or learning-based approaches.

\section{Proposed Approach}
We propose a novel approach, powered by DRL. Our framework adaptively sequences optimization modules based on input circuit structure and backend hardware characteristics. Its components include: (i) A DRL-based circuit rewriter that applies learned rewrite policies to reduce gate depth and two-qubit gate counts \cite{foselQuantumCircuitOptimization2021}. (ii) A domain-specific optimizer that performs logical numerical optimizations over gate parameters \cite{kuklianskyQFactorDomainSpecificOptimizer2023}. (iii) A parameterized circuit instantiator that improves compilation efficiency by solving for optimal template parameters during gate set translation \cite{younisQuantumCircuitOptimization2022a}. 

The orchestrator learns a coordination policy via DRL, using circuit-level features and backend-aware feedback (e.g., gate count, depth, fidelity) as state input. This enables adaptive selection of optimization sequences aligned with hardware constraints. Optimization is informed by the NISQ Analyzer and reward functions tailored to noise models. The result is a fully compiled circuit adapted for the target backend.

\begin{figure}
    \centering
    \includegraphics[width=0.46\textwidth]{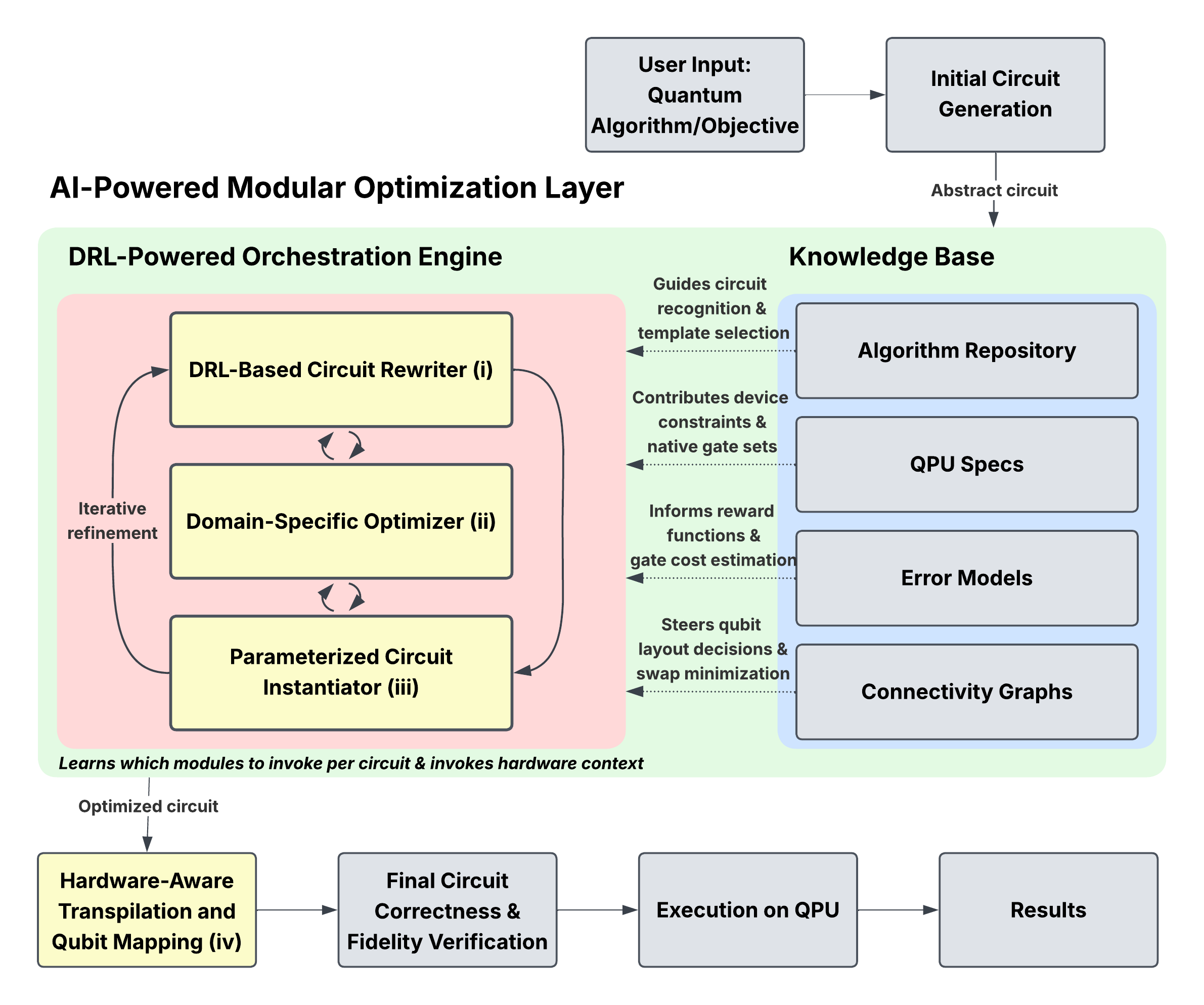}
    \caption{The OrQstrator framework architecture. (i), (ii), and (iii) are based on the works of Fösel et al. \cite{foselQuantumCircuitOptimization2021}, Kukliansky et al. \cite{kuklianskyQFactorDomainSpecificOptimizer2023}, as well as Younis and Iancu \cite{younisQuantumCircuitOptimization2022a}.}
    \label{fig:orqstrator-architecture}
\end{figure}

Figure \ref{fig:orqstrator-architecture} illustrates the overall architecture of our framework. A DRL-powered orchestration engine coordinates learned \cite{foselQuantumCircuitOptimization2021}, domain-specific \cite{kuklianskyQFactorDomainSpecificOptimizer2023}, and parameterized \cite{younisQuantumCircuitOptimization2022a} circuit optimizers using QPU-specific metadata from a shared knowledge base. The resulting circuit is hardware-aware \cite{dustdar_nisq_2020} and fidelity-optimized for NISQ execution. To our knowledge, no existing work orchestrated these methods in a modular, learned fashion targeting NISQ-era quantum devices. Building on the $\sim$27\% gate depth reduction from DRL \cite{foselQuantumCircuitOptimization2021} and $\sim$13\% gate count savings from instantiation \cite{younisQuantumCircuitOptimization2022a}, our goal is to exceed the combined reductions of individual methods, targeting $>$30\% overall gate/depth reduction while improving expected circuit fidelity. 




\section{Conclusion and Future Work}
We have presented OrQstrator, a novel approach that represents a step towards more intelligent quantum compilation. Our approach shifts quantum circuit optimization from rule-based approaches to learned strategy coordination. In the future, we will evaluate OrQstrator using a diverse suite of quantum circuits, including variational algorithms, such as the Variational Quantum Eigensolver (VQE) \cite{tillyVariationalQuantumEigensolver2022} and the Quantum Approximate Optimization Algorithm (QAOA) \cite{farhiQuantumApproximateOptimization2022}, arithmetic subroutines, and Clifford+T benchmarks \cite{onoratiRandomizedBenchmarkingIndividual2019}. Metrics include circuit depth, CX count, and simulated fidelity. Baselines include Qiskit's transpiler \cite{ibmTranspilerLatestVersion2025}, QFactor-only optimization, and DRL-only rewriters. All evaluations will target real device topologies (e.g., IBM Q \cite{IBMQuantumDocumentation}) and incorporate hardware-specific noise models.

\section*{Acknowledgment}
This work is funded by a grant (Q-Dev) from the Colorado Office of Economic Development and International Trade (OEDIT).

\bibliographystyle{ieeetr}
\bibliography{refs}
\end{document}